\documentclass[12pt]{article}
\usepackage{color}
\usepackage{enumerate,array}
\usepackage{amssymb,amsmath}

\newcommand{\B}{P_0,P_1}
\newcommand{\e}{\epsilon}

\makeatletter
\@addtoreset{equation}{section}
\makeatother
\begin{document}
\begin{flushright}
OU-HET 804
\end{flushright}
\begin{center}
\vspace{2mm}
{\LARGE The formulation of gauge-Higgs unification \vspace{2mm}

with dynamical boundary conditions}

\

{\large Kengo Yamamoto \vspace{2mm}\\
\emph{Department of Physics, Osaka University \\
Toyonaka 560-0043, Japan}}
\end{center}
\vspace{1cm}
\begin{abstract}
The boundary conditions on multiply connected extra dimensions play major rolls in gauge-Higgs unification theory. Different boundary conditions, having been given in ad hoc manner so far,  lead to different theories. To solve this arbitrariness problem of boundary conditions, we construct a formulation of gauge-Higgs unification with dynamics of boundary conditions on $M^4\times S^1/Z_2$. As a result, it is found that only highly restricted sets of boundary conditions, which lead to nontrivial symmetry breaking, practically contribute to the partition function. In particular, we show that for $SU(5)$ gauge group, sets of boundary conditions which lead to $SU(5)\to SU(3)\times SU(2)\times U(1)$ symmetry breaking are naturally included in the restricted sets. 
\end{abstract}
\thispagestyle{empty}

\newpage

\setcounter{page}{1}
\section{Introduction}
Gauge-Higgs unification(GHU) unifies gauge fields and Higgs scalar fields by considering gauge theory on higher dimensions \cite{phy, ann}. When multiply connected manifolds are introduced, dynamics of Wilson line phases lead to breakdown of gauge symmetry imposed on Lagrangian density. By using this Hosotani mechanism, GHU has been extensively investigated. There arise some difficulties for GHU due to introducing higher dimensions. One of them is the chiral fermion problem. One way to solve this problem is provided by considering GHU on orbifold. Furthermore, one can get natural solution for Higgs doublet-triplet mass splitting problem in $SU(5)$ grand unified theory (GUT) \cite{Rev, Kaw}. Also, the possibilities that one might achieve the unification of three families of quarks and leptons in higher-dimensional GUT on an orbifold have been proposed \cite{Bau, Fam}. But, in formulating GHU on orbifold, there remains subtlety that should be solved. At the moment, one imposes boundary conditions on multiply connected manifolds by hand, although there are a lot of possible boundary conditions imposed manifolds. We refer to this subtlety as arbitrariness problem of boundary conditions \cite{to}. This arbitrariness problem for GHU on orbifold was investigated  by N.Haba, M.Harada, Y.Hosotani, Y.Kawamura in detail \cite{nuc, pro}.  They classified equivalence classes for boundary conditions with using Hosotani mechanism, and analyzed their physics for each equivalence class. But, to solve this problem completely , we need dynamics of boundary conditions. Then, we must understand more fundamental theory to give this dynamics.

In this paper, we treat the boundary conditions as dynamical values, not those given by hand. For this goal, we have to generalize the present GHU formulation. We need the methods by which we can analyze systematically all possible configurations for the boundary conditions  in one framework. By using the matrix model analysis, we construct this framework. In this framework, we mainly focus on the natures of measures on integrations over the boundary conditions, to prove that only restricted sets of boundary conditions can contribute to the partition function, although we sum over all possible configurations for the boundary conditions. This restriction is common property in our formulation, irrespective of a detail of the action, and leads to the nontrivial gauge symmetry breaking. In particular, in the case of $SU(5)$ gauge group, the gauge-Higgs unification scenario is naturally restricted to only a few equivalence classes by the boundary conditions dynamics. Then, the equivalence class with standard model symmetry $SU(3)\times SU(2)\times U(1)$ as symmetry of boundary conditions is naturally included. 


In section 2, we give basic knowledge for GHU on orbfold, and classify each set of  boundary conditions to equivalence classes. In section 3, we give the formulation of GHU with dynamics of boundary conditions. In section 4, our formulation is applied to several examples. Section 5 is devoted to conclusions.

\section{Basic knowledges of GHU on $S^1/Z_2$}
\label{sec.1}
\subsection{Boundary conditions on $S^1/Z_2$}
\label{sub.1-1}
In this paper, we restrict our attention to GHU on $M^4\times S^1/Z_2$. The physics for this model was analyzed in Refs.\cite{nuc, pro}. $M^4$ is four-dimensional Minkowski spacetime. The fifth dimension $S^1/Z_2$ is obtained by identifying two points on $S^1$ by parity. Let $x$ and $y$ be coordinates of $M^4$ and $S^1$, respectively. $S^1$ has a radius R. In other words, a point $(x, y+2\pi R)$ is identified with a point $(x, y)$.  The orbifold $M^4\times S^1/Z_2$ is obtained by identifying $(x, y)\sim (x, y+2\pi R)\sim (x, -y)$. 

As a general principle the Lagrangian density has to be single-valued and gauge invariant on $M^4\times S^1/Z_2$. After a loop translation along $S^1$, each field needs to return to its original value only up to a global transformation of $U\in G$, where G is unitary gauge group imposed on Lagrangian density. It is called $S^1$ boundary condition. For gauge field $A_M \ \ \ (M=0\sim 3, 5)$
\begin{equation}
\label{eq.1-1-1}
A_M(x,y+2\pi R)=UA_M(x,y)U^{\dagger}.
\end{equation}
The $Z_2$-parity is specified by parity matrices. Around $y=0$
\begin{equation}
\label{eq.1-1-2}
\left(
\begin{array}{c}
A_{\mu}(x,-y) \\
A_{y}(x,-y)
\end{array}
\right)=P_0\left(
\begin{array}{c}
A_{\mu}(x,y) \\
-A_{y}(x,y)
\end{array}\right)P_0^{\dagger}
\end{equation}
and around $y=\pi R$
\begin{equation}
\label{eq.1-1-3}
\left(
\begin{array}{c}
A_{\mu}(x,\pi R-y) \\
A_{y}(x,\pi R-y)
\end{array}
\right)=P_1\left(
\begin{array}{c}
A_{\mu}(x,\pi R+y) \\
-A_{y}(x,\pi R+y)
\end{array}\right)P_1^{\dagger}.
\end{equation}
To preserve the gauge invariance, $A_y$ must have an opposite sign relative to $A_{\mu}$ under these transformations. As the repeated $Z_2$-parity operation brings a field configuration back to the original, $P_0$ must satisfy $P_0^2=1$. This means $P_0^{\dagger}=P_0$. $P_0$ must be an element of G up to an overall sign. This sign does not affect the result below so that we drop it in the following discussions. The same conditions apply to $P_1$, that is, 
\begin{equation}
\label{eq.1-1-4}
P_0^2=P_1^2=1
\end{equation}
Among $U, P_0$ and $P_1$, the relation 
\begin{equation}
\label{eq.1-1-5}
U=P_1P_0
\end{equation}
is satisfied. 

For scalar fields, the boundary conditions are specified by 
\begin{equation}
\label{eq.1-1-6}
\begin{array}{l}
\phi(x,-y)=\pm T_{\phi}[P_0]\phi(x,y)  \\
\phi(x,\pi R-y)=\pm e^{i\pi \beta_{\phi}}T_{\phi}[P_1]\phi(x,\pi R+y) \\
\phi(x,y+2\pi R)= e^{i\pi \beta_{\phi}}T_{\phi}[U]\phi(x,y) .
\end{array}
\end{equation}
$T_{\phi}[U]$ represents an appropriate representation matrix. The relation $T_{\phi}[U]=T_{\phi}[P_1]T_{\phi}[P_0]$ is also satisfied just as in (\ref{eq.1-1-5}). There are arbitrariness in the sign if the whole interaction terms in the Lagrangian remain invariant. $e^{i\pi 
\beta_{\phi}}$ must be either $+1$ or $-1$ due to $Z_2$-parity.

For Dirac fields, the boundary conditions are represented by
\begin{equation}
\label{eq.1-1-7}
\begin{array}{l}
\psi(x,-y)=\pm T_{\psi}[P_0]\gamma ^5\psi(x,y)  \\
	\psi(x,\pi R-y)=\pm e^{i\pi \beta_{\psi}}T_{\psi}[P_1]\gamma ^5\psi(x,\pi R+y) \\
\psi(x,y+2\pi R)= e^{i\pi \beta_{\psi}}T_{\psi}[U]\psi(x,y) .
\end{array}
\end{equation}
The phase factor $e^{i\pi \beta_{\psi}}$ must be either $+1$ or $-1$ just as for scalar fields. $(\gamma^{5})^2=1$ in our convention.

Therefore, the boundary conditions on $M^4\times S^1/Z_2$ are specified with $(P_0, P_1, U, \beta)$ and additional signs in (\ref{eq.1-1-6}) and (\ref{eq.1-1-7}). It is worthwhile to stress that the eigenvalues of $\B$ must be either $+1$ or $-1$ due to the condition $P_0^2=P_1^2=1$.

Next, we consider a gauge transformation on our system. Under a gauge transformation $\Omega(x,y)$, the fields change to 
\begin{equation}
\label{eq.1-1-8}
\begin{array}{c}
\displaystyle A_M(x,y)\to  A'_M(x,y)=\Omega(x,y)A_M(x,y)\Omega^{\dagger}(x,y)-\frac{i}{g}\Omega(x,y)\partial_M\Omega^{\dagger}(x,y) , \\
\phi(x,y) \to \phi'(x,y)=T_{\phi}[\Omega(x,y)]\phi , \ \ \ \ \ \ \ \psi(x,y)\to \psi'(x,y)=T_{\psi}[\Omega(x,y)]\psi.
\end{array}
\end{equation}
Generally, gauge transformations also change the given boundary conditions. After gauge transformation, the new fields $A_M'$ satisfy, instead of (\ref{eq.1-1-1}), (\ref{eq.1-1-2}) and (\ref{eq.1-1-3}),
\begin{equation}
\label{eq.1-1-9}
\displaystyle
\begin{array}{c}
\displaystyle
A'_M(x,y+2\pi R)=U'A'_M(x,y)U'^{ \dagger} -\frac{i}{g}U'\partial_MU'^{\dagger} \\ 
\left(
\begin{array}{c}
A'_{\mu}(x,-y) \\
A'_{y}(x,-y)
\end{array}
\right)
=P'_0\left(
\begin{array}{c}
A'_{\mu}(x,y) \\
-A'_{y}(x,y)
\end{array}
\right) 
P_0'^{ \dagger}-\displaystyle\frac{i}{g}P'_0
\left(
\begin{array}{c}
\partial_{\mu} \\
-\partial_{y}
\end{array}\right)P_0'^{\dagger} \\ 
\left(
\begin{array}{c}
A'_{\mu}(x,\pi R-y) \\
A'_{y}(x,\pi R-y)
\end{array}
\right)
=P'_1\left(
\begin{array}{c}
A'_{\mu}(x,\pi R+y) \\
-A'_{y}(x,\pi R+y)
\end{array}\right)P_1'^{\dagger}-\displaystyle\frac{i}{g}P'_1\left(
\begin{array}{c}
\partial_{\mu} \\
-\partial_{y}
\end{array}\right)P_1'^{\dagger}
\end{array}
\end{equation}
where, 
\begin{equation}
\label{eq.1-1-10}
\begin{array}{c}
U'=\Omega(x,y+2\pi R)U\Omega^{\dagger}(x,y) \\ 
P'_0=\Omega(x,-y)P_0\Omega^{\dagger}(x,y) \\
P'_1=\Omega(x,\pi R-y)P_1\Omega^{\dagger}(x,\pi R+y).
\end{array}
\end{equation}
Scalar  and fermion fields $\phi'$ and $\psi'$ satisfy relations similar to (\ref{eq.1-1-6}), (\ref{eq.1-1-7}), where $(\B, U)$ is replaced by $(P_0', P_1', U')$.

The gauge transformations which preserve the given boundary conditions are regard as the residual gauge invariance on the system. These transformations  which satisfy $U'=U$, $P_0'=P_0$ and $P_1'=P_1$ are defined by 
\begin{equation}
\label{eq.1-1-11}
\begin{array}{c}
\Omega(x,y+2\pi R)U=U\Omega(x,y) \\
\Omega(x,-y)P_0=P_0\Omega(x,y) \\
\Omega(x,\pi R-y)P_1=P_1\Omega(x,\pi R+y).
\end{array}
\end{equation}
(\ref{eq.1-1-11}) is called the symmetry of boundary conditions. Note that the physical symmetry can differ from the symmetry of boundary conditions.  When we consider the symmetry at low energies, namely gauge potential is independent on $y$: $\Omega=\Omega(x)$, the symmetry of boundary conditions is reduced to 
 \begin{equation}
\label{eq.1-1-12}
\Omega(x)U=U\Omega(x),\ \ \Omega(x)P_0=P_0\Omega(x),\ \ \Omega(x)P_1=P_1\Omega(x).
\end{equation}
That is, the symmetry is generated by generators which commute with $U$, $P_0$ and $P_1$.

We must regard theories with different boundary conditions as theories with different physical   content. But theories with different boundary conditions can be equivalent in physical content. If gauge transformation defined by $(\ref{eq.1-1-8})$ satisfies the conditions
\begin{equation}
\label{eq.1-1-13}
\partial_MP_0'=0,\ \ \partial_MP_1'=0, \ \ \partial_MU'=0,
\end{equation}
the two sets of boundary conditions are equivalent. We represent it as
\begin{equation}
\label{eq.1-1-14}
(P_0',P_1',U')\sim (P_0,P_1,U).
\end{equation}
The conditions (\ref{eq.1-1-13}) lead to $P_0'^{\dagger}=P_0', \ \ P_1'^{\dagger}=P_1'$. This $(P_0', P_1', U')$ also satisfy $(\ref{eq.1-1-4})$ $(\ref{eq.1-1-5})$, where $(\B, U)$ is replaced by $(P_0', P_1', U')$. The relation (\ref{eq.1-1-14}) defines equivalence classes, and the two theories in the same equivalence class lead to the same physical content although these theories may have different symmetries of boundary conditions. This equivalence of physical content is ensured by the Hosotani mechanism. This mechanism plays a major role in analyzing GHU. 
\subsection{Hosotani mechanism}
\label{sub.1-2}
The Hosotani mechanism that states theories in the same equivalence class lead to the same physical content takes place by the dynamics of Wilson line phases. The Hosotani mechanism in gauge theory defined on multiply connected manifolds is described by following statement \cite{ann}.
\begin{itemize}
\item[] \hspace*{1em}We give $WU$ defined by
\begin{equation}
\label{eq.1-2-1}
WU =\mathcal{P} \exp\bigg\{ig\int_C dy A_y\bigg\}U.
\end{equation}
The phases of $WU$ are called Wilson line phases. $U$ is the boundary condition of loop translation along non-contractible loop, $C$ is non-contractible loop, $\mathcal{P}$ denotes path ordered product. The eigenvalues of $WU$ are gauge invariant, so that these phases cannot be gauged away. Therefore, we should regard $WU$ as physical degrees of freedom. 
Wilson line phases are determined by dynamics of ($A_y^a, \ \frac{1}{2}\lambda^a \in \mathcal{H}_W$), where
\begin{equation}
\label{eq.1-2-2}
\mathcal{H}_W=\bigg\{\frac{\lambda^a}{2} ;\ \ \{\lambda^a,P_0\}=\{\lambda^a,P_1\}=0\bigg\}.
\end{equation}
That is, $\mathcal{H}_W$ is a set of generators which anti-commute with $\B$.

\hspace*{1em}Vacua of the system can degenerate at the classical level, but in general, the degeneracy of vacua is lifted by quantum effects. The vacuum given by the configuration of Wilson line phases, which minimizes the effective potential $V_{eff}$, becomes the physical vacuum of the system. 
If Wilson line phases have non-trivial configuration, the gauge symmetry imposed on system is spontaneously broken or restored by radiative corrections. As a result, gauge fields in lower dimension whose gauge symmetry is broken acquire masses from non-vanishing expectation values of the Wilson line phases. Some of matter fields also acquire masses. 

\hspace{1em}Two sets of boundary conditions which can be related by a boundary-condition-changing gauge transformation are physically equivalent, even if the symmetries of boundary conditions are different. This defines equivalence classes for boundary conditions. The physical symmetry of theory depends on the matter content of the theory through the expectation values of the Wilson line phases. One can determine the physical symmetry of theory by the combination of boundary conditions and the expectation values of the Wilson line phases. 
\end{itemize}

We can determine physical symmetry of the theory under given boundary conditions $(P_0, P_1, U)$, by using this Hosotani mechanism. We suppose $V_{eff}$ is minimized by constant $\langle A_y\rangle$, and $\exp (i2\pi gR\langle A_y\rangle) \ne I$. Then, $\langle A_y\rangle$ is transformed to $\langle A'_y\rangle=0$ by gauge potential $\Omega(x,y)=\exp\{ig(y+\alpha)\langle A_y\rangle\}$. After this transformation, boundary conditions change to 
\begin{equation}
\label{eq.1-2-3}
(P_0^{sym},P_1^{sym},U^{sym},\beta)\equiv (e^{2ig\alpha \langle A_y\rangle}P_0,\ e^{2ig(\alpha+\pi R)\langle A_y\rangle}P_1,\ WU,\ \beta).
\end{equation}
As only extra dimensional components $A_y$ whose generators anti-commutate with $P_0,P_1$ can have non-vanishing expectation values,  the boundary conditions (\ref{eq.1-2-3}) indeed satisfy (\ref{eq.1-1-4}), (\ref{eq.1-1-5}). As $\langle A'_y\rangle=0$ in this gauge, physical symmetry of the theory agrees with the symmetry of boundary conditions. Then, physical symmetry of theory is determined by 
\begin{equation}
\label{eq.1-2-4}
H^{sym}=\bigg\{\frac{\lambda^a}{2};\ \ [\lambda^a,P_0^{sym}]=[\lambda^a,P_1^{sym}]=0\bigg\}.
\end{equation}
\subsection{Classification of equivalence classes}
\label{sub.1-3}
In this subsection, we will classify the equivalence classes for boundary conditions in $SU(N)$ gauge theory by using $SU(2)$ subgroup gauge transformations \cite{pro}. The matrices $\B$ may not be diagonal in general. We can always diagonalize one of them, say $P_0$, through a global gauge transformation, but $P_1$ might not be diagonal. However, in Ref.\cite{pro}, we know each equivalence class has ($\B$) that are both diagonal representations. So, let us consider diagonal $\B$, which are specified by three non-negative integers $(p, q, r)$ such that
\begin{equation}
\label{eq.1-3-1}
\begin{array}{l}
\mathtt{diag}\ P_0=\overbrace{(+1,\cdots , +1,+1,\cdots , +1,-1,\cdots ,-1,-1,\cdots ,-1)}^N \\
\mathtt{diag}\ P_1=\underbrace{(+1,\cdots ,+1,}_p\underbrace{-1,\cdots , -1,}_q\underbrace{+1,\cdots , +1,}_r\underbrace{-1,\cdots ,-1)}_{s=N-p-q-r},
\end{array}
\end{equation}
where $N\ge p,q,r,s\ge 0$. We denote the boundary conditions indicated $(p,q,r)$ as $[p;q,r;s]$. The matrix $P_0$ is interchanged with $P_1$ by the interchange of $q$ and $r$. To illustrate the boundary-changing local gauge transformations, we consider an $SU(2)$ gauge theory with $(P_0,P_1,U)=(\tau_3,\tau_3,I)$. After gauge transformation $\Omega=\exp\big\{i\big(\frac{\alpha y}{2\pi R}\big)\tau_2\big\}$, we obtain the equivalence relation 
\begin{equation}
\label{eq.1-3-2}
(\tau_3,\tau_3,I)\sim (\tau_3,e^{i\alpha \tau_2}\tau_3,e^{i\alpha \tau_2}).
\end{equation}
In particular, for $\alpha=\pi$ we have 
\begin{equation}
\label{eq.1-3-3}
(\tau_3,\tau_3,I)\sim (\tau_3,-\tau_3,-I).
\end{equation}
Using this equivalence relation, we can have the following equivalence relations in $SU(N)$ gauge theory:
\begin{equation}
\label{eq.1-3-4}
\begin{array}{r}
[p,q,r,s]  \sim [p-1;q+1,r+1;s-1] \ \ \ \mathtt{for}\ \ p,s\ge 1 \\ 
 \sim [p+1;q-1,r-1;s+1] \ \ \ \mathtt{for}\ \ q,r\ge 1
\end{array}
\end{equation}
The sets of boundary conditions connected by this equivalence relations  lead to the same physical content. We can completely classify the equivalence classes in $SU(N)$ gauge theory on orbfold, by using (\ref{eq.1-3-1}), (\ref{eq.1-3-4}). It has been showed that the number of equivalence classes in $SU(N)$ gauge theory on orbfold equals to $(N+1)^2$ \cite{pro}.
\section{Reformulation of gauge-Higgs unification with dynamical boundary conditions}
\label{sec.2}
In this section, we will give a formulation for GHU including the dynamics of boundary conditions, and show only restricted sets of boundary conditions practically contribute to the partition function.
\subsection{Definition of model}
\label{sub.2-1}
The partition function for $SU(N)$ GHU on orbfold is given by: 
\begin{equation}
\label{eq.2-1-1}
\mathrm{Z}=\int_C dP_0 \int_C dP_1 \int  \mathcal{D}A_M\mathcal{D}\bar{\psi}\mathcal{D}\psi \bigg|_{P_0,P_1} e^{iS(A_M,\psi,P_0,P_1)},
\end{equation}
where,
\begin{equation}
\label{eq.2-1-2}
C=\{P_i\in U(N),\ \ P_i^2=1\}\ \ \ i=1,2
\end{equation}
and $S(A_M,\psi , P_0,P_1)$ is the action depending on gauge fields, fermion fields and boundary conditions. We suppose that the action $S(A_M, \psi, P_0, P_1)$ is invariant under gauge transformation on fields $A_M, \psi$, but the boundary conditions may not be so. The symbol $|_{P_0,P_1}$ means we restrict functional integral regions for fields $A_M, \psi$ to preserve the boundary conditions.  $dP_0,dP_1$ are defined as $U(N)$ invariant measures.
\subsection{Natures of integral with $dP_0,dP_1$ }
\label{sub.2-2}
We will discuss general natures of integration over the boundary conditions $\int_C dP_0\int_C dP_1$. First, we consider the following transformation for integral variable $P_0$
\begin{equation}
\label{eq.2-2-1}
P_0=U^{\dagger}P_0'U,
\end{equation}
where $U\in  U(N)$. Under this transformation, the integration over  ${P_0}$ converts into 
\begin{equation}
\begin{array}{c}
\label{eq.2-2-2}
\int_C dP_0 =\int_{C'} d[U^{\dagger}P_0'U] \\
C\equiv \{P_0\in U(N),\ \ P_0^2=1\} \\
C'=UCU^{\dagger}.
\end{array}
\end{equation}
Note $d[U^{\dagger}P_0'U]=dP_0'$ from the property of invariant measure. Since $(P'_0)^2=1,\ \ P_0'\in C'$, we can see $C=C'$. So, we find
\begin{equation}
\label{eq.2-2-3}
\int_C dP_0=\int_CdP_0'.
\end{equation}
The same discussion can apply to $P_1$.

Next, we will give the method which splits integration of a function depending on $\B$ between diagonal variables and off-diagonal variables \cite{adv, math}. We start with 
\begin{equation}
\label{eq.2-2-4}
\mathrm{F}=\int_CdP_0\int_CdP_1 f(\B),
\end{equation}
where $f(\B)$ is a function depending on $\B$, and we assume $f(\B)$ is invariant under transformation $P_0\to UP_0U^{\dagger},\ \ P_1 \to UP_1U^{\dagger}\ \ \ \ \ U\in U(N)$. That is, 
\begin{equation}
\label{eq.2-2-5}
f(UP_0U^{\dagger},UP_1U^{\dagger})=f(\B).
\end{equation}
Then, we define the following function
\begin{equation}
\label{eq.2-2-6}
\begin{array}{c}
\Delta^{-1}(P_0)\equiv \displaystyle \int dU\prod_{1\le i <j \le N}\delta^{(2)}[(UP_0U^{\dagger})_{ij}]  \\
\delta^{(2)}[(UP_0U^{\dagger})_{ij}] \equiv \delta[\Re (UP_0U^{\dagger})_{ij}]\delta[\Im (UP_0U^{\dagger})_{ij}].
\end{array}
\end{equation}
 Here, $ dU$ is the invariant measure of $U(N)$. Substituting the function defined by (\ref{eq.2-2-6}) to (\ref{eq.2-2-4}), we find 
\begin{equation}
\label{eq.2-2-8}
\begin{array}{c}
\mathrm{F}=\displaystyle \int_C dP_0 \int_C dP_1 \Delta(P_0)\int dU\prod_{1\le i <j \le N}\delta^{(2)}[(UP_0U^{\dagger})_{ij}] f(\B) .
\end{array}
\end{equation}
Change the variable as $P_0 =U^{\dagger}P_0'U$. Since the function (\ref{eq.2-2-6}) is invariant under this transformation, and by using (\ref{eq.2-2-3}), we find
\begin{equation}
\label{eq.2-2-11}
\mathrm{F}=\int dU\int_C dP_0' \int_C dP_1 \Delta(P_0')\prod_{1\le i <j \le N}\delta^{(2)}[(P_0')_{ij}] f(U^{\dagger}P_0'U,P_1) .
\end{equation}
Change the variable as $P_1=U^{\dagger}P_1'U$, and using (\ref{eq.2-2-3}) where $P_0$ is replaced with $P_1$ and (\ref{eq.2-2-5}), (\ref{eq.2-2-11}) equals to 
\begin{equation}
\label{eq.2-2-12}
\mathrm{F}=\int dU\int_C dP_0' \int_C dP_1' \Delta(P_0')\prod_{1\le i <j \le N}\delta^{(2)}[(P_0')_{ij}] f(P_0',P_1') .
\end{equation}
We normalize $\int dU=1$. Next, we change the integral region by regularization parameter $\mu$ to regularize (\ref{eq.2-2-12}).
\begin{equation}
\label{eq.2-2-13}
C\to \hat{C}\equiv\{P_0\in U(N),\  \rho_i=\pm e^{i\mu_i}, \ 0\le \mu_i \le \mu\ll 1\}\ \ \ \ \mu:real .
\end{equation}
$\rho_i\ (1\le i\le N)$ denote eigenvalues of $P_0$. This change preserves the relation (\ref{eq.2-2-3}), and in the limit $\mu\to 0$ we can restore it to the original definition. At the end of our calculation, we must take the limit $\mu\to 0$. Carry out  integration of $P_0'$ with $\delta$ function in (\ref{eq.2-2-11}), it becomes 
\begin{equation}
\label{eq.2-2-14}
\mathrm{F}=\int d\Lambda_0\int_C dP_1' \Delta(\Lambda_0)f(\Lambda_0, P_1'),
\end{equation} 
where 
\begin{equation}
\label{eq.2-2-15}
\Delta^{-1}(\Lambda_0)=\frac{(2\pi)^N}{\displaystyle\prod_{1\le i<j \le N}|\epsilon_i-\epsilon_je^{i\mu_{ij}}|^2}, \ \ \ \mu_{ij}=\mu_j-\mu_i .
\end{equation}
$\epsilon_i, \epsilon_j$ are $+1$ or $-1$.

The symbol $\int d\Lambda_0$ denotes integration over only diagonal matrices in the integral region $\hat{C}$. In the regularization (\ref{eq.2-2-13}), it is represented by
\begin{equation}
\label{eq.2-2-16}
\int d\Lambda_0=\displaystyle \sum_{\pm 1}\int_0^{\mu}\prod_{1\le n\le N}d\mu_n.
\end{equation}
$\displaystyle\sum_{\pm 1}$ means the summation over all combinations we assign $+1$ or $-1$ to $\e_i \ (1\le i\le N)$ in (\ref{eq.2-2-14}), (\ref{eq.2-2-15}). 

We can apply the same calculation and regularization from (\ref{eq.2-2-6}) to (\ref{eq.2-2-15}) for $P_1$, and we have
\begin{equation}
\label{eq.2-2-17}
\mathrm{F}= \int d\Lambda_0\int d{\Lambda_1} \Delta(\Lambda_0)\Delta(\Lambda_1)\int dU f(\Lambda_0, U^{\dagger}\Lambda_1U),
\end{equation}
where, 
\begin{equation}
\label{eq.2-2-18}
\Delta^{-1}(\Lambda_1)=\frac{(2\pi)^N}{\displaystyle\prod_{1\le p<q \le N}|\epsilon_p'-\epsilon_q'e^{i\mu_{pq}'}|^2},\ \ \ \mu_{pq}'=\mu_q'-\mu_p'
\end{equation}
\begin{equation}
\label{eq.2-2-19}
\int d\Lambda_1=\displaystyle\sum_{\pm 1}\int^{\mu'}_0 \prod_{1\le m\le N}d\mu_m'
\end{equation}
$\mu' \ll 1$ is the regularization parameter, and $\e'_p, \e'_q$ are $+1$ or $-1$.  

For boundary conditions $P_0, P_1 \in U(N),\ \ N\ge 3$, taking the limit $\mu, \mu' \to 0$ in (\ref{eq.2-2-17}) lead to $\mathrm{F}\to 0$. It means the integral regions for the boundary conditions correspond to the regions of measure zero in the $U(N)$ invariant measure in $U(N)$ group manifold. We must renormalize the partition function (\ref{eq.2-1-1}) to make it well-defined.
\subsection{Integration of partition function for boundary conditions }
\label{sub.2-3}
In this subsection, we will apply the method discussed in subsection \ref{sub.2-2} to the model defined in subsection \ref{sub.2-1}. We will find that only some of sets of boundary conditions practically contribute to the partition function. First, as noted in the end of subsection \ref{sub.2-2} we need to divide the partition function (\ref{eq.2-1-1}) by the volume $\int_C dP_0\int_C dP_1$. We regularize the integral $\int_C dP_0\int_C dP_1$ in the denominator and numerator with parameters $\mu, \mu'$, and adopt the following normalization when we take the limit,
\begin{equation}
\label{eq.2-3-22}
\frac{\int_{\hat{C}}dP_0\int_{\hat{C}}dP_1}{\int_{\hat{C}}dP_0\int_{\hat{C}}dP_1}\to 1,\ \ \ \ \ \mu,\mu'\to 0.
\end{equation}
 According to the discussion in subsection \ref{sub.2-2}, this volume can be written as  
\begin{equation}
\label{eq.2-3-20}
V \equiv\int_{\hat{C}} dP_0\int_{\hat{C}} dP_1 =\int d\Lambda_0\int d\Lambda_1 \Delta(\Lambda_0)\Delta(\Lambda_1)  .
\end{equation}
The notations follow the definitions in subsection \ref{sub.2-2}. The normalized partition function, $\mathrm{Z}$, is defined by 
\begin{equation}
\label{eq.2-3-21}
\mathrm{Z}=V^{-1}\int_{\hat{C}} dP_0 \int_{\hat{C}} dP_1 \int  \mathcal{D}A_M\mathcal{D}\bar{\psi}\mathcal{D}\psi \bigg|_{P_0,P_1} e^{iS(A_M,\psi,P_0,P_1)}.
\end{equation}
The field values are not defined in this regularization since $P_0^2, P_1^2 \ne 1$. We redefine parity transformation matrices $\hat{P_0}, \hat{P_1}$ as
 \begin{equation}
 \label{eq.2-3-24}
 \hat{P_0}\equiv (P_0^{-2})^{\frac{1}{2}}P_0, \ \ \ \hat{P_1}\equiv (P_1^{-2})^{\frac{1}{2}}P_1,
 \end{equation}
 where 
 \begin{equation}
 \label{eq.2-3-25}
 A^{\frac{1}{2}}=U\Lambda^{\frac{1}{2}}U^{\dagger}, \ \ 
 \Lambda^{\frac{1}{2}}=\left(
 \begin{array}{ccc}
 \sqrt{a_1} & &\\
  &\sqrt{a_2} & \\
  & & \ddots
 \end{array}\right) \ \ \ A\in U(N).
 \end{equation}
 $a_i\ \ (i=1, 2, \cdots)$ are the eigenvalues of $A$, and we choose the positive square root of the eigenvalues as the convention. In this prescription, we find the eigenvalues of $\hat{P_0}, \hat{P_1}$ are $+1$ or $-1$, and $\hat{P_0}^2=\hat{P_1}^2=1$. We can restore those to the original definitions in the limit $\mu, \mu'\to 0$. The integrand of the boundary conditions is well-defined function. From now on, the symbol $|_{\B}$ means we restrict the functional integral regions for fields $A_M, \psi$ to preserve the boundary conditions $\hat{P_0}, \hat{P_1}$. 
 
The next step is to divide the integration in the partition function into diagonal components and off-diagonal components of boundary condition matrices $\B$, according to subsection \ref{sub.2-2}.
\begin{equation}
\label{eq.2-3-1}
\begin{aligned}
\mathrm{Z}=V^{-1}\int_{\hat{C}} dP_0 \int_{\hat{C}} dP_1 \int  \mathcal{D}A_M\mathcal{D}\bar{\psi}\mathcal{D}\psi \bigg|_{P_0,P_1} \Delta(P_0) \int dU \delta^{(2)}(UP_0U^{\dagger}) e^{iS(A_M,\psi,P_0,P_1)}
\end{aligned},
\end{equation}
where,
\begin{equation}
\label{eq.2-3-2}
\delta^{(2)}(UP_0U^{\dagger})\equiv \displaystyle \prod_{1\le i<j\le N}\delta^{(2)}[(UP_0U^{\dagger})_{ij}].
\end{equation}
In (\ref{eq.2-3-1}), we change the integration variable from $P_0$ to $P'_0=UP_0U^{\dagger}$, and using (\ref{eq.2-2-3}), we find 
\begin{equation}
\label{eq.2-3-3}
\begin{array}{r}
\mathrm{Z}=V^{-1}\int dU \int_{\hat{C}} dP_0' \int_{\hat{C}} dP_1 \int  \mathcal{D}A_M\mathcal{D}\bar{\psi}\mathcal{D}\psi \bigg|_{U^{\dagger}P_0'U,P_1} \Delta(P_0')\delta^{(2)}(P_0') \\
\times e^{iS(A_M,\psi,U^{\dagger}P_0'U,P_1)} .
\end{array}
\end{equation}
Following the discussion in subsection \ref{sub.2-2}, we integrate out $\int dP_0'$ in (\ref{eq.2-3-3}) with $\delta^{(2)}(P_0')$. (\ref{eq.2-3-3}) equals to 
\begin{equation}
\label{eq.2-3-4}
\begin{array}{r}
\mathrm{Z}=\displaystyle V^{-1}\int dU \int d\Lambda_0 \int_{\hat{C}} dP_1\Delta(\Lambda_0) \int  \mathcal{D}A_M\mathcal{D}\bar{\psi}\mathcal{D}\psi \bigg|_{U^{\dagger}\Lambda_0U ,P_1} \\
\times e^{iS(A_M,\psi,U^{\dagger}\Lambda_0U,P_1)} .
\end{array}
\end{equation}
$U^{\dagger}\Lambda_0 U$ is the unitary transformation for $U\in U(N)$. But we can regard this transformation as the unitary transformation for $U'\in SU(N)$. One can multiply this transformation by diagonal $U(N)$ element $\Lambda$, as $U^{\dagger}\Lambda_0U$ preserves its value. That is, $U'^{\dagger}\Lambda_0 U'=U^{\dagger}\Lambda_0 U$ for $U'=\Lambda U$. So, by multiplying $U$ by suitable $\Lambda$, we can find $U'=\Lambda U\ \ U'\in SU(N)$ for arbitrary $U \in U(N)$. Therefore, we can rewrite (\ref{eq.2-3-4}) as 
\begin{equation}
\label{eq.2-3-5}
\begin{array}{r}
\mathrm{Z}=\displaystyle V^{-1}\int dU \int d\Lambda_0 \int_{\hat{C}} dP_1\Delta(\Lambda_0) \int  \mathcal{D}A_M\mathcal{D}\bar{\psi}\mathcal{D}\psi \bigg|_{U'^{\dagger}\Lambda_0U' ,P_1} \\
\times e^{iS(A_M,\psi,U'^{\dagger}\Lambda_0U',P_1)} \\
U'\in SU(N).
\end{array}
\end{equation}
Change the integration variable from $P_1$ to $P'_1=U'P_1U'^{\dagger}$ and use (\ref{eq.2-2-3}) where $P_0$ is replaced with $P_1$, We have
\begin{equation}
\label{eq.2-3-6}
\begin{array}{r}
\mathrm{Z}=\displaystyle V^{-1}\int dU \int d\Lambda_0 \int_{\hat{C}} dP_1' \Delta(\Lambda_0) \int  \mathcal{D}A_M\mathcal{D}\bar{\psi}\mathcal{D}\psi \bigg|_{U'^{\dagger}\Lambda_0U',U'^{\dagger}P_1'U'}\\
\times e^{iS(A_M,\psi,U'^{\dagger}\Lambda_0U',U'^{\dagger}P_1'U')} .
\end{array}
\end{equation}
(\ref{eq.2-3-6}) equals to the original system that has the boundary conditions $(\hat{\Lambda}_0, \hat{P_1'})$ up to the global gauge transformation $U'$. The system should be independent on global gauge. So, the relation (\ref{eq.2-3-6}) becomes
\begin{equation}
\label{eq.2-3-7}
\begin{array}{r}
\mathrm{Z}=\displaystyle V^{-1}\int dU \int d\Lambda_0 \int_{\hat{C}} dP_1' \Delta(\Lambda_0) \int  \mathcal{D}A_M\mathcal{D}\bar{\psi}\mathcal{D}\psi \bigg|_{\Lambda_0,P_1'}  \\
\times e^{iS(A_M,\psi,\Lambda_0,P_1')} .
\end{array}
\end{equation}
Normalize $\int dU=1$, and apply in the same procedure to $P_1'$. Then, equation (\ref{eq.2-3-7}) becomes
\begin{equation}
\label{eq.2-3-10}
\begin{array}{l}
\mathrm{Z}=\frac{\displaystyle \sum_{\pm 1}\int^{\mu}_0 \prod_{1\le n\le N}d\mu_n\int^{\mu'}_0\prod_{1\le m\le N}d\mu_m' \Delta(\Lambda_0)\Delta(\Lambda_1)I(A_M,\psi , \Lambda_0, \Lambda_1)}{\displaystyle\sum_{\pm 1}\int^{\mu}_0 \prod_{1\le n'\le N}d\mu_{n'}\int^{\mu'}_0\prod_{1\le m'\le N}d\mu_{m'}' \Delta(\Lambda_0)\Delta(\Lambda_1)}
\end{array}
\end{equation}
where
\begin{equation}
\label{eq.2-3-31}
I(A_M,\psi , \Lambda_0, \Lambda_1)\equiv \int dU\int  \mathcal{D}A_M\mathcal{D}\bar{\psi}\mathcal{D}\psi \bigg|_{\Lambda_0,U^{\dagger}\Lambda_1U} e^{iS(A_M,\psi,\Lambda_0,U^{\dagger}\Lambda_1U)} .
\end{equation}
We suppose $I(A_M,\psi , \Lambda_0, \Lambda_1)$ is almost constant function on the integral variables $\mu_n$ and $\mu_m'$, compared with $\Delta(\Lambda_0), \Delta(\Lambda_1)$. Then, we can replace the function $\displaystyle \int^{\mu}_0 \prod_{1\le n\le N}d\mu_n\int^{\mu'}_0\prod_{1\le m\le N}d\mu_m' \Delta(\Lambda_0)\Delta(\Lambda_1)$ with the integrand on particular values $\mu_n, \mu_m'\ \ (0<\mu_n, \mu_m'<\mu)$ times the integral regions by mean-value theorem. We can put the conditions $\mu_{ij}\ne 0, \mu_{pq}'\ne 0, \ \ (1\le i,p <j,q\le N)$ if $\e_i$ and $\e_j$ or $\e_p'$ and $\e_q'$ have the same sign, since these values correspond to the maximum or minimum of the integrand. After this replacement, the integral regions of $d\mu_n, d\mu_m'$ between the denominator and numerator in (\ref{eq.2-3-10}) cancel out. As a result, we have
\begin{equation}
\label{eq.2-3-32}
\mathrm{Z}=\frac{\displaystyle \sum_{\pm 1} \prod_{1\le i,p<j,q\le N}|\e_i-\e_je^{i\mu_{ij}}|^2|\e_p'-\e_q'e^{i\mu_{pq}'}|^2I(A_M,\psi , \Lambda_0, \Lambda_1)}{\displaystyle\sum_{\pm 1}\prod_{1\le k,v<l,w\le N}|\e_k-\e_le^{i\mu_{kl}}|^2|\e_v'-\e_w'e^{i\mu_{vw}'}|^2}.
\end{equation}
In the summation $\displaystyle \sum_{\pm 1}$ , the factors $|1-e^{i\mu_{ij}}|^2$ and $|1-e^{i\mu_{pq}'}|$ give 0 to each term in (\ref{eq.2-3-31}) when we take the limit $\mu, \mu' \to 0$. We suppose "$a$" is the lowest number of the factors ,such as $|1-e^{i\mu_{ij}}|$, each term has in (\ref{eq.2-3-32}). We can find the lowest number of the factors such as $|1-e^{i\mu_{pq}'}|$ is also $a$. Then, we multiply the denominator and numerator in (\ref{eq.2-3-32}) by $|1-e^{i\mu}|^{-2a}$. As taking the limit $\mu\to 0$, we can see
\begin{equation}
\label{eq.2-3-33}
\frac{|1-e^{i\mu_{ij}}|}{|1-e^{i\mu}|}=\bigg|\frac{\mu_{ij}}{\mu}\bigg|\to C_{ij} >0.
\end{equation}
$C_{ij}$ must be finite value in order to be consistent with mean-value theorem. There is at least one term which has finite value in denominator and numerator of (\ref{eq.2-3-32}) in this limit. Such finite terms correspond to the terms which have the highest number of pairs of different signs substituted for $\e_i\ \ (1\le i\le N)$.  The other terms go to $0$ when we take the limit $\mu\to 0$. Since the $\e_i\ \ (1\le i\le N)$ denote the eigenvalues of $P_0$, we can find only sets of the eigenvalues of $P_0$ which have the highest number of pairs of different signs contribute to the partition function in (\ref{eq.2-3-32}). We will have the same conclusion if previous discussion is applied to the integral of $P_1$. 

Relating these results to the discussion about the dimensions of unitary conjugate class for a particular set of eigenvalues gives us more observations. In $U(N)$ group if a set of eigenvalues has no the identical eigenvalue, the submanifold which consists of the elements of unitary conjugate class for its set of eigenvalues has the highest dimensions among the submanifolds of the unitary conjugate classes. And the more identical eigenvalues a set of eigenvalues includes, the less dimensions the submanifold of its unitary conjugate class has \cite{cla}. Therefore, in our case, unitary conjugate classes which includes the highest number of the pairs $+1$, $-1$ as the eigenvalues of $\B$ have the highest dimensions among the sets of  eigenvalues of $\B$, and only these boundary conditions contribute to the partition function in the integral process.

Next, let us consider the case that boundary conditions ($\Lambda_0', \Lambda_1'$) are related to the diagonal boundary conditions ($\Lambda_0, \Lambda_1$) by permutation of the eigenvalues sets. We will show $I(A_M, \psi, \Lambda_0', \Lambda_1')$ gives a identical contribution to the partition function as $I(A_M, \psi, \Lambda_0, \Lambda_1)$. Since $(\Lambda_0', \Lambda_1')$ is the permutation of eigenvalues sets in $(\Lambda_0,\Lambda_1)$, it satisfies the relations 
\begin{equation}
\label{eq.2-3-14}
\begin{array}{c}
\Lambda_0'=V_0^{\dagger}\Lambda_0V_0 \ \ \ \ \ \Lambda_1'=V_1^{\dagger}\Lambda_1V_1\\
V_0,V_1 \in SU(N),
\end{array} 
\end{equation}
and the factors $\displaystyle \prod_{1\le i<j\le N}|\e_i-\e_je^{i\mu_{ij}}|^2$, $\displaystyle \prod_{1\le p<q\le N}|\e_p'-\e_q'e^{i\mu_{pq}'}|^2$ give identical contribution to $I(A_M, \psi, \Lambda_0', \Lambda_1')$ and $I(A_M, \psi, \Lambda_0, \Lambda_1)$. One find for the boundary conditions $(\Lambda_0',\Lambda_1' )$ in (\ref{eq.2-3-31})
\begin{equation}
\label{eq.2-3-16}
\begin{array}{l}
I(A_M,\psi , \Lambda_0', \Lambda_1')  \\
=\int dU\int  \mathcal{D}A_M\mathcal{D}\bar{\psi}\mathcal{D}\psi \bigg|_{\Lambda_0',U^{\dagger}\Lambda_1'U} e^{iS(A_M,\psi,\Lambda_0',U^{\dagger}\Lambda_1'U)} \\
=\int dU\int  \mathcal{D}A_M\mathcal{D}\bar{\psi}\mathcal{D}\psi \bigg|_{V^{\dagger}_0\Lambda_0V_0,U^{\dagger}V_1^{\dagger}\Lambda_1V_1U} e^{iS(A_M,\psi,V_0^{\dagger}\Lambda_0V_0,U^{\dagger}V_1^{\dagger}\Lambda_1V_1U)} .
\end{array}
\end{equation}
Under global gauge transformation $\Lambda_0'\to V_0\Lambda_0'V_0^{\dagger},\ \ \ U^{\dagger}\Lambda_1'U\to V_0U^{\dagger}\Lambda_1' UV_0^{\dagger}$, we find
\begin{equation}
\label{eq.2-3-17}
\begin{array}{l}
I(A_M,\psi , \Lambda_0', \Lambda_1')  \\
=\int dU\int  \mathcal{D}A_M\mathcal{D}\bar{\psi}\mathcal{D}\psi \bigg|_{\Lambda_0,V_0U^{\dagger}V_1^{\dagger}\Lambda_1V_1UV_0^{\dagger}} e^{iS(A_M,\psi,\Lambda_0,V_0U^{\dagger}V_1^{\dagger}\Lambda_1V_1UV_0^{\dagger})}. 
\end{array}
\end{equation}
Using the property of  $\int dU$ invariant measure, we have
\begin{equation}
\label{eq.2-3-18}
\begin{array}{l}
I(A_M,\psi , \Lambda_0', \Lambda_1') =I(A_M,\psi , \Lambda_0, \Lambda_1).
\end{array}
\end{equation}
Then, we can see in (\ref{eq.2-3-18}) $I(A_M, \psi, \Lambda_0, \Lambda_1)$ and $I(A_M, \psi, \Lambda_0', \Lambda_1')$ give the identical contributions to (\ref{eq.2-3-10}). 
According to the discussion in subsection \ref{sub.1-3}, there is the gauge transformation which relates the boundary conditions ($\Lambda_0,\Lambda_1$) to ($\Lambda_0',\Lambda_1'$). Then, It is worthwhile to state the boundary conditions ($\Lambda_0,\Lambda_1$) and ($\Lambda_0',\Lambda_1'$) are in the same equivalence class. According to the discussion of Appendix A in Ref.\cite{pro}, we can see there is at least one both diagonal boundary conditions in each equivalence class. Then, on the process that arbitrary boundary conditions change to both diagonal representations by global and local gauge transformations, there is no transformation which changes the eigenvalues set of the boundary conditions. So, arbitrary boundary conditions $(\B)$ and its eigenvalue set $(\Lambda_0, \Lambda_1)$ belong to the same equivalence class. Since a permutation $(\Lambda_0', \Lambda_1')$ of diagonal representations $(\Lambda_0, \Lambda_1)$ belong to the equivalence class with $(\Lambda_0, \Lambda_1)$, we conclude equivalence classes for GHU on $S^1/Z_2$ in $SU(N)$ gauge theory are completely classified by eigenvalues sets for boundary conditions. Therefore, on the process that we compute some physical observables, the integrand on $\int dU$ in (\ref{eq.2-3-10}) is independent of the variable $U$, so it is sufficient to compute only about the both diagonal representations $(P_0,P_1)=(\Lambda_0,\Lambda_1)$ if we want to know some physical observables.
\section{Application to several examples}
\label{sec.3}
In this section, we apply the formulation we had in section \ref{sec.2} to $SU(2), SU(3)$, $SU(5)$ gauge theory. In particular, We are interested in $SU(5)$ gauge theory as the candidate for GUT. As a consequence of the boundary conditions dynamics presented here, sets of the boundary conditions will be highly restricted. In $SU(5)$ case, we will show these restricted sets include the equivalence classes which have the standard model symmetry $SU(3)\times SU(2)\times U(1)$ as the symmetry of boundary conditions.

First, we consider $SU(2)$ gauge theory on $M^4\times S^1/Z_2$ as the simplest example. In the case, there is only one equivalence class of boundary conditions that gives a non-vanishing contribution to the partition function. It is characterized by the eigenvalue set
\begin{equation}
\label{eq.4-10}
\left\{
\begin{aligned}
P_0=\{+1,-1\}\\
P_1=\{+1,-1\}
\end{aligned}\right\}
\end{equation}
This boundary conditions lead to the symmetry breaking $SU(2)\to U(1)$ as  symmetry of boundary conditions. 

The next example is $SU(3)$ gauge theory. In this case, the four sets of boundary conditions and their equivalence classes contribute to the partition function. These equivalence classes are characterized by following eigenvalue sets
 \begin{equation}
\label{eq.4-11}
\begin{array}{ll}
\begin{aligned}
&(1)\\
&\left\{
\begin{aligned}
P_0=\{+1,+1,-1\}\\
P_1=\{+1,+1,-1\}
\end{aligned}\right\}\end{aligned}&\begin{aligned}
&(2)\\
&\left\{
\begin{aligned}
P_0=\{+1,+1,-1\}\\
P_1=\{+1,-1,-1\}
\end{aligned}\right\}\end{aligned} \\
\begin{aligned}
&(3)\\
&\left\{
\begin{aligned}
P_0=\{+1,-1,-1\}\\
P_1=\{+1,+1,-1\}
\end{aligned}\right\}\end{aligned}&\begin{aligned}
&(4)\\
&\left\{
\begin{aligned}
P_0=\{+1,-1,-1\}\\
P_1=\{+1,-1,-1\}
\end{aligned}\right\}.\end{aligned}
\end{array}
\end{equation}
The boundary conditions (1) and (4) lead to the symmetry breaking $SU(3)\to SU(2)\times U(1)$. On the other hand, the boundary conditions (2) and (3) lead to the symmetry breaking $SU(3)\to U(1)\times U(1)$. The partition function in (\ref{eq.2-3-32}) is written as
\begin{equation}
\label{eq.4-2}
\mathrm{Z}=C_1\mathrm{I}_{(1)}+C_2\mathrm{I}_{(2)}+C_3\mathrm{I}_{(3)}+C_4\mathrm{I}_{(4)}.
\end{equation}
Here, $I_{(i)}\ \ i=1\sim 4$ indicate the $I(A_M,\psi , \Lambda_0, \Lambda_1)$ in (\ref{eq.2-3-31}). The subscript indices mean we substitute corresponding boundary conditions (i) in (\ref{eq.4-11}) into $I_{(i)}$. Since the factors $\displaystyle \prod_{1\le i<j\le N}|\e_i-\e_j'e^{i\mu_{ij}}|^2$, $\displaystyle \prod_{1\le p<q\le N}|\e_p'-\e_q'e^{i\mu_{pq}'}|^2$ give the overall constant in (\ref{eq.4-2}), we dropped this constant. $C_i$ denote the coefficients corresponding to all permutation in the boundary conditions (i). In $SU(3)$ case, these constants are 
\begin{equation}
\label{eq.4-20}
C_i=(_3C_1)^2 \ \ \ \ \ i=1\sim 4.
\end{equation}
So, we can see all coefficients are the same, and drop this coefficients as overall constants.

Finally, we investigate $SU(5)$ gauge theory example. Just as in the $SU(3)$ example, four boundary conditions sets and their equivalence classes contribute to the partition function. These equivalence classes are characterized by
\begin{equation}
\label{eq.4-1}
\begin{array}{ll}
\begin{aligned}
&(1)\\
&\left\{
\begin{aligned}
P_0=\{+1,+1,+1,-1,-1\}\\
P_1=\{+1,+1,+1,-1,-1\}
\end{aligned}\right\}\end{aligned}&\begin{aligned}
&(2)\\
&\left\{
\begin{aligned}
P_0=\{+1,+1,+1,-1,-1\}\\
P_1=\{+1,+1,-1,-1,-1\}
\end{aligned}\right\}\end{aligned} \\
\begin{aligned}
&(3)\\
&\left\{
\begin{aligned}
P_0=\{+1,+1,-1,-1,-1\}\\
P_1=\{+1,+1,+1,-1,-1\}
\end{aligned}\right\}\end{aligned}&\begin{aligned}
&(4)\\
&\left\{
\begin{aligned}
P_0=\{+1,+1,-1,-1,-1\}\\
P_1=\{+1,+1,-1,-1,-1\}
\end{aligned}\right\}.\end{aligned}
\end{array}
\end{equation}
Boundary conditions (2) and (3) lead to the symmetry breaking $SU(5)\to SU(2)\times SU(2)\times U(1)\times U(1)$. We should mention the boundary conditions (1) and (4) have $SU(3)\times SU(2)\times U(1)$ standard model symmetry as the symmetry of boundary conditions.  The partition function consists of the four part that correspond to the boundary conditions $(1)\sim (4)$ respectively. We note that physical symmetry depends on the matter content. 
\section{Conclusion}
In this paper, we have supposed that fundamental theory can describe the dynamics of the boundary conditions in GUH, and have discussed the natures of the measures $dP_0$, $dP_1$. In the present scenario of GHU, the orbfold boundary conditions are imposed in an ad hoc manner among many possible choices. The boundary conditions can be classified in equivalence classes by using the Hosotani mechanism. Two theories in the same equivalence class lead to the identical physical content. In particular, the number of equivalence classes of $SU(N)$ gauge theory on $M^4\times S^1/Z_2$ is $(N+1)^2$. In other words, $SU(N)$ gauge theory on $M^4\times S^1/Z_2$ has $(N+1)^2$ different theories. 

We have showed only the boundary conditions which have the highest number of the pair $+1$, $-1$ in eigenvalues of $\B$ eventually contribute to the partition function in our formulation. The submanifolds which consist of  these boundary conditions as the elements have the highest dimensions among submanifolds of the equivalence classes for boundary conditions. In $SU(N)$ gauge theory where $N$ is odd, the four equivalence classes practically contribute to partition function. These equivalence classes lead to nontrivial breakdown of the symmetries imposed on Lagrangian density. To determine which set of boundary conditions is selected as physical state in these four sets of boundary conditions, we need to evaluate the effective potentials for each set of boundary conditions. But the difference between two equivalence classes may appear to be infinite. It is known that the energy difference become finite in supersymmetric GHU.

To consider the arbitrariness problem completely, we should regard GHU as an effective theory given by the more fundamental theory. The fundamental theory may select the lowest energy state as physical state by giving the dynamics of the boundary conditions. 
\section*{Acknowledgements}
I would like to give a special thanks to Y.Hosotani for helpful discussions and his supports. And I would like to thank S.Yamaguchi, K.Hashimoto and H.Hatanaka for their comments and encouragement. I could complete this work owing to their supports.

\end{document}